\documentclass[epj,showpacs,lengthcheck]{svjour}
\usepackage{epsfig}
\usepackage{amsbsy}
\usepackage{amssymb}
\usepackage{amsmath}
\usepackage{amsbsy}
\usepackage{latexsym}
\usepackage{graphics}

\begin{document}

\title{Collective Excitations of Harmonically Trapped Ideal Gases}
\author{Bert Van Schaeybroeck \and Achilleas Lazarides}

%
\institute{Instituut voor Theoretische Fysica,\\
Katholieke Universiteit Leuven,\\ Celestijnenlaan 200 D,\\ B-3001
Leuven, Belgium}
\date{\today}

\abstract{We theoretically study the collective excitations of an
ideal gas confined in an isotropic harmonic trap. We give an exact
solution to the Boltzmann-Vlasov equation; as expected for a
single-component system, the associated mode frequencies are
integer multiples of the trapping frequency. We show that the
expressions found by the scaling ansatz method are a special case
of our solution. Our findings, however, are most useful in case
the trap contains more than one phase: we demonstrate how to
obtain the oscillation frequencies in case an interface is present
between the ideal gas and a different phase.}  \maketitle

\section{Introduction}\label{intro}
Since the experimental realization of quantum degenerate gases,
the role of particle interactions has been thoroughly studied.
While weak interactions characterize most ultracold systems due to
their diluteness, the use of Feshbach resonances allows the study
of very strongly interacting systems. Although interactions are
generally important, many experimentally realized gases may be
well-modelled by ideal gases. For example,
collisions are suppressed in fully-polarized Fermi gases at
temperatures well below their Fermi temperature because s-wave
interactions are forbidden due to Pauli blocking. Such gases
appear for instance in the experiments on imbalanced fermion
gases~\cite{zwierlein,partridge,lazarides} and in phase-segregated
Bose-Fermi mixtures~\cite{ospelkaus,zaccanti,vanschaeybroeck2}. A
description in terms of ideal gases is also appropriate in
ultracold two-component Fermi systems when the interspecies
scattering length is tuned to zero by means of a Feshbach
resonance~\cite{du}. Moreover, it is well-known that sufficiently
dilute gases become collisionless; as a result, collisionless
behavior also sets in as the temperature of trapped gases well
above the temperature of quantum degeneracy~\cite{pethick}.

For a hydrodynamic gas (or superfluid) under experimentally relevant
conditions, the variables relevant for studying the collective excitations
are, for example, the local density and velocity; on the other hand, for a
collisionless fluid, one must, in general, find the entire distribution
function in momentum- and coordinate-space. In some cases, one may extract
dynamical solutions using a scaled form of the equilibrium distribution
function~\cite{kagan,castin,bijlsma,guery,pedri}. However, the scaling
approach fails for a two-component system with an
interface~\cite{lazarides,vanschaeybroeck2}. Other methods to obtain the
collective mode frequencies include the method of averaging, the sum-rule
approach, and the random-phase
approximation~\cite{guery2,amoruso,vichi,bruun,capuzzi,santamore,maruyama}.

We present here a solution for the Boltzmann-Vlasov equation of an
ideal gas in an isotropic harmonic trap. For the single-component
system, we recover the spectrum $n\omega_0$ with $n$ integer and
$\omega_0$ the trapping frequency. We then show how our solution
may be used to study a two-component system with an interface by
writing down the appropriate boundary conditions at the interface.
For the monopole modes, we give analytical expressions which may
be straightforwardly used to calculate their frequencies in
isotropic and highly-elongated elliptical traps. Finally, we
discuss damping of these modes.

\section{Boltzmann-Vlasov Equation}\label{sec:1}
An ideal gas, confined by an isotropic potential
$m\omega_0^2\mathbf{r}^2/2$, is described by a distribution
function $f(\mathbf{r},\mathbf{p},t)$, evolving according to the
Boltzmann-Vlasov equation:
\begin{equation}\label{vlasov}
    m\partial_t f+\mathbf{p}\cdot\boldsymbol{\nabla}_\mathbf{r}
    f-m^2\omega_0^2\mathbf{r}\cdot\boldsymbol{\nabla}_\mathbf{p}f=0.
\end{equation}
In the following $f_0$ denotes the equilibrium solution, for which
$\partial_t f_0=0$. Collective oscillations may be studied by
looking at periodic deviations from $f_0$. Writing $f=f_0+\nu
e^{-i\omega t}$ we obtain the result
\begin{equation}\label{solution}
    \nu=\mathcal{C}\left(\left\{L_j,E_j\right\}\right)
    \left[\sum_{j=x,y,z}a_j(p_j+i m\omega_0
    r_j)^2\right]^{\omega/2\omega_0},
\end{equation}
with the $a_j$ arbitrary dimensionless constants and $\mathcal{C}$
an arbitrary function of the angular momentum components
$L_j=(\mathbf{r}\times\mathbf{p})_j$ and of the
$E_j=(p^2_j/m+m\omega_0^2r_j^2)/2$ (with $j=x,y,z$). Note that
solution~\eqref{solution} is not obtained by linearization so
$\nu$ may be large compared to $f_0$.


\section{Single-component System}\label{sec:2}
Consider a trap containing a single particle component. Due to the
absence of boundary conditions, the function $\mathcal{C}$ in
Eq.~\eqref{solution} is arbitrary; physically, $\nu/\mathcal{C}$
must then be continuous everywhere. Hence, $\omega$ has to be
chosen such that no discontinuities appear when the coordinates
$r_j$ cross zero. This is guaranteed if $\omega=n\omega_0$ (from now on,
$n$ is an integer), which is the well-known spectrum of a
single-component trapped collisionless gas~\cite{pethick}. We show
now that indeed all experimentally relevant collective modes can
be obtained from Eq.~\eqref{solution}.

The lowest dipole or Kohn mode can be found by taking, for example,
$\mathbf{a}=(0,0,1)$; it is then clear that $\omega=\omega_0$
ensures that $\nu/\mathcal{C}$ converges to the same value if
$r_z\rightarrow 0^+$ or $r_z\rightarrow 0^-$ when $p_z<0$. It is
also easy to see that, when $\mathcal{C}$ is rotationally
invariant in coordinate space, the density deviation $\delta\rho
\propto\int \mathrm{d}^3 \mathbf{p}\, \nu$ is simply proportional
to $r_z$, which, as expected for the dipole mode, is proportional
to the spherical harmonic $Y_{\ell=1}^0$.

Due to rotational symmetry in space the monopole modes can be
described by taking $\mathbf{a}=(1,1,1)$ such that
\begin{equation*}
\nu=\mathcal{C}\left(\mathbf{L}^2,E\right)\left[\mathbf{p}^2-m^2\omega_0^2\mathbf{r}^2+
2im\omega_0\mathbf{r}\cdot\mathbf{p}\right]^{\omega/2\omega_0},
\end{equation*}
with $E=\sum_j E_j$. Arguments similar to those for the dipole
mode lead to the conclusion that $\omega=2n\omega_0$, which is the
well-known spectrum for monopole modes. Moreover, for the lowest
excitation ($n=1$), the density deviation $\delta\rho$ is
proportional to the spherical harmonic $Y_{\ell=0}^0$.

Finally, for the quadrupole mode, two of the $a_j$ must be equal
while the third opposite in sign. Choosing $\mathbf{a}=(1,1,-2)$,
one finds that $\omega=2\omega_0$ and that $\delta\rho\propto
x^2+y^2-2z^2\propto Y_{\ell=2}^{0}$, as it must\footnote{Note that
for this to be true, it is necessary that $\mathcal{C}$ is
rotationally symmetric.}.

A highly-elongated elliptical trap may be modelled by an isotropic
2D confinement $U(\mathbf{r})=m\omega_0^2(r_y^2+r_z^2)/2$. This
results in a cylindrical configuration with the $x$-axis along the
axial direction and rotational symmetry in the $y-z$ plane. The
angular dependencies of $\delta\rho$ for the lowest monopole,
dipole and quadrupole modes are known to vary as
$e^{i\ell\theta}\propto (y+iz)^\ell$ with $\theta$ the angle in
the $y-z$ plane and $\ell=0,1,2$. The Kohn mode ($\ell=1$) can
therefore be recovered by an appropriate superposition of
Eq.~\eqref{solution} with $\mathbf{a}=(0,1,0)$ and
Eq.~\eqref{solution} with $\mathbf{a}=(0,0,1)$, together with a
rotational-invariant function $\mathcal{C}$. In an analogous
manner, and using a $\mathcal{C}$-function which depends on $L_x$,
one may obtain the quadrupole mode ($\ell=2$).

We now explain the relation between Eq.~\eqref{solution} and the
scaling ansatz method, widely used in the
literature~\cite{kagan,castin,bijlsma,guery}.

\section{Relation to the Scaling Ansatz}\label{sec:2}

Consider a confined single-component gas with equilibrium
distribution $f_0$. Due to the radial symmetry, $f_0$ can be
written as a function of $\mathbf{L}^2$ and $E$ only; we assume that
$f_0$ is a function of $E$ only. According to the scaling ansatz,
the distribution function $f$ of a trapped single-component system
at a certain time $t$ may be written in terms of the
equilibrium distribution $f_0$ as follows:
\begin{equation}
    f(\mathbf{r},\mathbf{p},t)=f_0(\mathbf{r}(t),\mathbf{p}(t)).
\end{equation}
In order to find the monopole and the quadrupole modes one
continues by takes the scaling functions $r_j(t)=r_j/\alpha_j$ and
$p_j(t)=\alpha_j p_j-m\dot{\alpha}_jr_j$ where the dot denotes the
derivative with respect to time and $\alpha_j$ is a periodic
function of time~\cite{kagan,castin,bijlsma,guery}. For small
deviations from equilibrium, one writes $\alpha_j=1+\varepsilon
a_je^{-i\omega t}$ with $\varepsilon\ll 1$ and $a_j$ dimensionless
constants. Linearization of $f$ in terms of $\varepsilon$ yields:
\begin{align}\label{scaling1}
&f=f_0+\varepsilon e^{-i\omega t} m^{-1}(\partial_{_E}f_0)\\
&\times\left(\sum_{j=x,y,z}\left[a_j(p_j+i
m\omega_0r_j)^2+ima_j(\omega-2\omega_0)r_jp_j\right]\right).\nonumber
\end{align}
Clearly $f$ reduces to Eq.~\eqref{solution} when
$\omega=2\omega_0$, which, as we have seen, is indeed the
frequency associated with the lowest monopole and quadrupole modes
of an ideal gas.

The Kohn mode, on the other hand, may be recovered by substitution
of $r_z(t)=r_z-\beta$ and $p_z(t)=p_z-\dot{\beta}$ while for
$j=x,\,y$, one takes $r_j(t)=r_j$ and $p_j(t)=p_j$. Assuming a
time-dependent function $\beta(t)=\varepsilon e^{-i\omega t}$ and
linearizing in terms of $\varepsilon$ yields:
\begin{equation}\label{scaling2}
f= f_0+i\varepsilon e^{-i\omega t}(\partial_{_E} f_0)\left[\omega
p_z+im\omega_0^2r_z\right].
\end{equation}
Again this reduces to our solution~\eqref{solution} in case
$\mathbf{a}=(0,0,1)$, and $\omega=\omega_0$.

We therefore conclude that, in a linearized form, the expressions
for the distribution functions used by the scaling ansatz method
reduce to our solution~\eqref{solution} in case $\omega=\omega_0$
and $\omega=2\omega_0$. Note that, while our
solution~\eqref{solution} solves the Boltzmann-Vlasov equation,
Eqs.~\eqref{scaling1} and \eqref{scaling2} do not, except for
specific values of $\omega$. Therefore, the reason why, for a
single-component system, the correct spectrum arises from
Eq.~\eqref{solution}, and from Eqs.~\eqref{scaling1} and
\eqref{scaling2}, is different. Note further that, although the
scaling ansatz does not give the most general solution for the
Boltzmann-Vlasov equation, it may be generalized to find
oscillation frequencies using a Boltzmann equation which includes
a mean-field and collision term~\cite{kagan,castin,bijlsma,guery}.

\section{Two-Component System with an Interface}
Assume in the following the presence of two phases, separated in
the trap by an interface at radial position $\zeta$. We
investigate the boundary conditions at the interface for the
ideal-gas phase and leave the nature of the other phase
unspecified. Experimentally relevant cases include the imbalanced
fermion gases~\cite{zwierlein,partridge,lazarides,vanschaeybroeck}
and phase-segregated Bose-Fermi
mixtures~\cite{ospelkaus,zaccanti,vanschaeybroeck2,modugno,ferlaino2}.

Under the assumption that particles are
specularly reflected, conservation of energy and particle number
relate the distribution function of incoming and outgoing
particles at the interface~\cite{bekarevich}:
\begin{equation}\label{bkbc}
    \nu(\zeta,\chi)-\nu(\zeta,-\chi)=2m
    \chi\dot{\zeta}(\partial_{_E}f_0),
\end{equation}
with $\chi$ defined by $\mathbf{r}\cdot\mathbf{p}=\chi rp$ where
$r=|\mathbf{r}|$ and $p=|\mathbf{p}|$.

Consider the point $\mathbf{r}=(0,0,\zeta)$ on the interface. The
sole result of a particle reflecting there is its change of
velocity component $p_z$ to $-p_z$. Therefore, aside from $L_z$,
$E_x$ and $E_y$, also $L_x$, $L_y$ and $E_z$ are invariant under
reflection, such that $\mathcal{C}(\chi)=\mathcal{C}(-\chi)$. The
solution of Eq.~\eqref{solution} satisfying the boundary
condition~\eqref{bkbc} at position $\mathbf{r}=(0,0,\zeta)$ is
then:
\begin{equation}\label{bkbc2}
\nu=2m
\chi\dot{\zeta}(\partial_{_E}f_0)\,[1-e^{-i\omega\tau}]^{-1},
\end{equation}
where,
\begin{equation}\label{tau}
\tau=\omega_0^{-1}\mathrm{Arg}\left[a_x p_x^2+a_y p_y^2+a_z(\chi
p+im\omega_0 \zeta)^2 \right].
\end{equation}
A general solution for $\nu$ at $\mathbf{r}=(0,0,\zeta)$ can be
obtained by taking superpositions of Eq.~\eqref{bkbc2} with each
different values of $\mathbf{a}$. In the following, we show how
one can study breathing modes for gases confined in isotropic and
highly-elongated harmonic traps. Multipole modes may then be
studied in an analogous manner.

\textit{Spherical trap} --- The breathing modes can be found by
imposing the condition of rotational symmetry in Eq.~\eqref{tau},
such that:
\begin{equation}\label{tau2}
    \tau=\omega_0^{-1}\mathrm{Arg}\left[p^2- m^2\omega_0^2
     \zeta^2+2im\omega_0\chi\zeta p\right].
\end{equation}
Physically, $\tau$ is the time for a particle departing radially
from the interface, to return to it. For a fully-degenerate Fermi
gas, the values for $\zeta$ and $\omega$ for which damping occurs
are depicted in Fig.2 of Ref.~\cite{vanschaeybroeck2}. Note that,
due to the rotational invariance Eq.~\eqref{bkbc2} is valid at any
point on the interface.

\textit{Highly Elongated Traps} --- In case of highly-elongated
harmonic traps, one can approximate the trapping potential by
$U(\mathbf{r})=m\omega_0(r_y^2+r_z^2)/2$. The breathing modes can
be studied assuming $\mathbf{a}=(1,1,0)$ and considering $\nu$ in
the spatial point $(0,0,\zeta)$ where Eq.~\eqref{bkbc2} is valid
with:
\begin{align}\label{tau4}
\tau=\omega_0^{-1}\mathrm{Arg}&\left[p^2(\chi^2+(1-\chi^2)\cos^2\phi)\right.\nonumber\\
&\quad\quad\left.-m^2\omega_0^2 \zeta^2+2im\omega_0\chi \zeta
p\right].
\end{align}
Here $\phi$ is the azimuth of $\mathbf{p}$ in the $x$-$y$-$z$
coordinate system.

Apart from the boundary condition~\eqref{bkbc}, which by itself
ensures that the mean velocity of the collisionless gas at the
interface is equal to $\dot{\zeta}$, another boundary condition
applies, ensuring local mechanical equilibrium at the interface.
This condition is the Laplace equation and it relates the radial
pressure tensors $\Pi_{rr}$ of the inner phase $A$ and the outer
phase $B$ at the interface:
\begin{align}\label{laplace}
\Pi_{rr}^A-\Pi_{rr}^B=\gamma_{AB}\left(1/R_{_1}+1/R_{_2}\right).
\end{align}
Here $R_{_1}$ and $R_{_2}$ are the principal radii of curvature of
the interface and $\gamma_{AB}$ is the interface tension of phase
$A$ and $B$.

Knowledge of the distribution function $\nu$ from
Eq.~\eqref{tau} allows a direct calculation of the pressure tensor
$\Pi_{rr}$ in Eq.~\eqref{laplace} for the ideal-gas phase.
Assuming one knows the pressure tensor for the other phase, all
the necessary ingredients are present to find the collective mode
frequencies, as shown in Refs.~\cite{lazarides,vanschaeybroeck}.

\section{Work Done by a Moving Interface}
We now show that the total energy transfer through the interface
over one oscillation period vanishes unless a pole is present in
the pressure tensor, in which case the collective mode gets
damped. We determine also a criterion for this damping to occur.

The work done by a moving interface onto a collisionless gas is
most simply expressed using the pressure tensor normal to that
interface, $\Pi_{rr}$, and the average velocity of the gas in the
same direction, $u_r$ (here $r$ denotes the radial coordinate). At
$\mathbf{r}=(0,0,\zeta)$, the functions $\delta\Pi_{rr}$ and $u_r$
vary as:
\begin{align*}
    \delta\Pi_{rr}\propto e^{-i\omega t}\int_0^\infty
        \mathrm{d}p \int_{0}^1
        \mathrm{d}\chi\,\chi^3
        p^4(\partial_{_E} f_0)\text{cot}\left(\frac{\omega\tau}{2}\right)
\end{align*}
and
\begin{align*}
u_r&\propto ie^{-i\omega t}\int_0^\infty \mathrm{d}p\,
p^3(\partial_{_E} f_0),
\end{align*}
where we have used that $\tau(\chi)=-\tau(-\chi)$. The work done per unit area
during one oscillation period $T=2\pi/\Re(\omega)$ is given by
\begin{equation*}
    \Delta W=\int_{0}^{T}\mathrm{d}t\, \Re\left(\delta\Pi_{rr}\right)
    \Re\left(u_r\right).
\end{equation*}
If there appears no pole in the integrand of $\delta\Pi_{rr}$,
then $\delta\Pi_{rr}$ and $u_r$ are $\pi$ out of phase and the
total work done during a single cycle averages to zero. If, on the
other hand, the integrand of $\delta\Pi_{rr}$ has a pole for some
$\chi$, then one must pass below that pole when integrating over
$\chi$~\cite{landau}; as a result, $\delta\Pi_{rr}$ acquires an
imaginary part and is no longer $\pi$ out of phase with $u_r$.
This results in net work being done on the collisionless gas
during one period. In other words, it corresponds to damping of the
collective modes, with their energy transferred to single-particle
modes in the collisionless gas (in analogy to Landau damping).

One may ask now when a pole appears in the integrand of
$\delta\Pi_{rr}$. From Eq.~\eqref{bkbc2},
$\nu(\zeta,\chi)$ diverges if:
\begin{equation}\label{dampingcondition}
    \tau(\zeta,\chi)=2\pi n/\omega,
\end{equation}
with $n$ again an integer. This gives rise to the following
criterion: damping at the interface can only happen when the
collective mode frequency exceeds $2\omega_0$. Indeed, since
$|\tau|\leq\pi/\omega_0$, Eq.~\eqref{dampingcondition} can only be
satisfied when $\omega\geq 2\omega_0$

Finally, note that it can be shown that, if no pole appears in the
integrand of $\delta\Pi_{rr}$ at the interface, then there is no
pole anywhere in the trap.

\section{Conclusions}
We have presented a solution (see Eq.~\eqref{solution}) for the
Boltz\-mann-Vlasov equation, describing the collective excitations
of an ideal gas in an isotropic harmonic trap. If the trap is
entirely filled by the ideal gas, the associated mode frequencies
are simply integer multiples of the trapping frequency. In that
case, we proved that the expressions for the distribution function
used by the scaling ansatz method, are a special case of our
solution. We further applied our solution to a trap consisting of
an ideal gas in contact with a different phase by means of an
interface. Taking the particles of the ideal-gas phase to
specularly reflect on the interface, we show how the breathing
mode frequencies may be obtained in isotropic and highly-elongated
elliptical traps. Finally, we discuss how, in the presence of an
interface, damping may arise.

\section{Acknowledgement}
We acknowledge partial support by Project No.~FWO G.0115.06;
A.L.~is supported by Project No.~GOA/2004/02 and B.V.S. by the
Research Fund K.U.Leuven.

\end{document}